\documentclass[12pt,leqno]{article}

\title{Jeans type analysis of chemotactic collapse}

\usepackage{amsthm,amsmath,amssymb,a4wide,psfig,epsf}
\usepackage{color}
\def\mb#1{\setbox0=\hbox{$#1$}\kern-.025em\copy0\kern-\wd0
\kern-0.05em\copy0\kern-\wd0\kern-.025em\raise.0233em\box0}

\begin{document}

\author{Pierre-Henri Chavanis and Cl\'ement Sire}
\maketitle
\begin{center}
Laboratoire de Physique Th\'eorique (IRSAMC, CNRS), Universit\'e Paul
Sabatier,\\ 118, route de Narbonne, 31062 Toulouse Cedex, France\\
E-mail: {\it chavanis{@}irsamc.ups-tlse.fr \&
clement.sire{@}irsamc.ups-tlse.fr }

\vspace{0.5cm}
\end{center}

\begin{abstract}

We perform a linear dynamical stability analysis of a general
hydrodynamic model of chemotactic aggregation [Chavanis \& Sire,
Physica A, {\bf 384}, 199 (2007)]. Specifically, we study the
stability of an infinite and homogeneous distribution of cells against
``chemotactic collapse''. We discuss the analogy between the
chemotactic collapse of biological populations and the gravitational
collapse (Jeans instability) of self-gravitating systems. Our
hydrodynamic model involves a pressure force which can take into
account several effects like anomalous diffusion or the fact that the
organisms cannot interpenetrate. We also take into account the
degradation of the chemical which leads to a shielding of the
interaction like for a Yukawa potential. Finally, our
hydrodynamic model involves a friction force which quantifies the
importance of inertial effects. In the strong friction limit, we
obtain a generalized Keller-Segel model similar to the generalized
Smoluchowski-Poisson system describing self-gravitating Langevin
particles. For small frictions, we obtain a hydrodynamic model of
chemotaxis similar to the Euler-Poisson system describing a
self-gravitating barotropic gas. We show that an infinite and
homogeneous distribution of cells is unstable against chemotactic
collapse when the ``velocity of sound'' in the medium is smaller than
a critical value. We study in detail the linear development of the
instability and determine the range of unstable wavelengths, the
growth rate of the unstable modes and the damping rate, or the
pulsation frequency, of the stable modes as a function of the friction
parameter and shielding length. For specific equations of state, we
express the stability criterion in terms of the density of cells.

\vskip0.5cm


\end{abstract}

\section{Introduction}
\label{intro}

In biology, many microscopic organisms (bacteria, amoebae, endothelial
cells,...) or even social insects (like ants) interact through the
phenomenon of chemotaxis \cite{murray}. These organisms deposit a
chemical (pheromone, smell, food,...) that has an attractive
\footnote{The case of repulsive chemotaxis due to a ``poison'' can
also be of interest and will be considered in a future contribution.} 
effect on the organisms themselves. Therefore, in addition to their
diffusive motion, they move preferentially along the gradient of
concentration of the chemical they secrete (chemotactic flux). When
chemotactic attraction prevails over diffusion, this process can lead
to a ``chemotactic collapse'' (see \cite{horstmann} for a review)
resulting in the aggregation of the organisms.  In this way, some
structures can form like clusters (clumps) or even network patterns
(filaments). Therefore, the chemotactic interaction can explain
several features of the morphogenesis of biological colonies.  The
chemotactic aggregation of biological populations is usually described
in terms of the Keller-Segel model
\cite{ks}. This is a parabolic model consisting in two coupled
differential equations.  The first equation is a drift-diffusion
equation describing the evolution of the concentration of cells and
the second equation is a reaction-diffusion equation with terms of source and
degradation describing the evolution of the concentration of the
secreted chemical. This model ignores inertial effects and assumes
that the drift velocity of the organisms is directly induced by a
chemotactic ``force'' proportional to the concentration gradient of
the chemical. The Keller-Segel model can reproduce the formation of
clusters (clumps) by chemotactic collapse [4-19]. This reflects
experiments on bacteria like {\it Escherichia coli} or amoebae like
{\it Dictyostelium disco\"ideum} exhibiting pointwise concentration
\cite{ks}.  However, parabolic models fail at describing the formation
of network patterns (filaments). These filaments are observed in
experiments of capillary blood vessels formation
\cite{carmeliet}. They correspond to the spontaneous self-organization
of endothelial cells during vasculogenesis, a process occuring during
embryologic development. In order to account for these structures,
more general models of chemotaxis have been introduced
\cite{gamba,filbet,csbio}. They have the form of hydrodynamic
(hyperbolic) models taking into account inertial effects. These models
can reproduce the formation of filaments that are interpreted as the
beginning of a vasculature.  This phenomenon is responsible of
angiogenesis, a major factor for the growth of tumors
\cite{chaplain}. Interestingly, these filaments share some analogies
with the large-scale structures in the universe that are described by
similar hydrodynamic equations
\cite{peebles,vergassola}.

Recently, we have introduced a general kinetic model
of chemotactic aggregation based on generalized stochastic
processes, non linear mean field Fokker-Planck equations and
generalized thermodynamics \cite{csbio}. From these kinetic equations, we have
derived a hydrodynamic model of the form
\begin{equation}
\label{intro1} {\partial\rho\over\partial t}+\nabla\cdot (\rho{\bf u})=0,
\end{equation}
\begin{equation}
\label{intro2} \frac{\partial {\bf u}}{\partial t}+({\bf u}\cdot
\nabla){\bf u}=-{1\over\rho}\nabla p+\nabla c-\xi {\bf u},
\end{equation}
\begin{equation}
\label{intro3}{\partial c\over\partial t}=D_{c}\Delta c-k c+h\rho.
\end{equation}
It involves a pressure force $-\nabla p$, where $p=p(\rho)$ is a
barotropic equation of state that can take into account several
effects like anomalous diffusion or the fact that the particles do not
interpenetrate. It also involves a friction force $-\xi {\bf u}$ which
measures the importance of inertial effects. For $\xi=0$, we recover
the hyperbolic model introduced by Gamba {\it et al.}
\cite{gamba}. For $\xi\rightarrow +\infty$, we can neglect the
inertial term in the momentum equation (\ref{intro2}) leading to $\xi
{\bf u}\simeq-{1\over\rho}\nabla p+\nabla c$ (overdamped
limit). Substituting this relation in the equation of continuity
(\ref{intro1}), we obtain the generalized Keller-Segel model
\begin{equation}
{\partial\rho\over\partial t}=\nabla \cdot \left \lbrack\chi \left (\nabla p-\rho\nabla c\right )\right \rbrack, \label{intro4}
\end{equation}
\begin{equation}
\label{intro5}{\partial c\over\partial t}=D_{c}\Delta c-k c+h\rho,
\end{equation}
where $\chi=1/\xi$.  Interestingly, this model of chemotaxis is
similar to a model of self-gravitating Langevin particles
\cite{virial} described by the damped Euler-Poisson system
\begin{equation}
\label{intro6} {\partial\rho\over\partial t}+\nabla\cdot (\rho{\bf u})=0,
\end{equation}
\begin{equation}
\label{intro7} \frac{\partial {\bf u}}{\partial t}+({\bf u}\cdot
\nabla){\bf u}=-{1\over\rho}\nabla p-\nabla \Phi-\xi {\bf u},
\end{equation}
\begin{equation}
\label{intro8}\Delta\Phi=S_{d}G\rho.
\end{equation}
For $\xi=0$, it reduces to the barotropic Euler-Poisson system
\cite{bt} and for $\xi\rightarrow +\infty$, we obtain the generalized
Smoluchowski-Poisson system
\begin{equation}
{\partial\rho\over\partial t}=\nabla \cdot \left \lbrack \frac{1}{\xi} \left (\nabla p+\rho\nabla \Phi\right )\right \rbrack, \label{intro9}
\end{equation}
\begin{equation}
\label{intro10}\Delta\Phi=S_{d}G\rho.
\end{equation}
In this analogy, we see that the concentration $-c({\bf r},t)$ of the
chemical plays the same role as the gravitational potential $\Phi({\bf
r},t)$. In biology, the interaction is mediated by a {\it material}
substance (the secreted chemical) while the physical interpretation of
the gravitational potential in astrophysics is more abstract
\footnote{The notion of ``force at distance'' in the Newtonian
theory has been criticized at several occasions in the history of
physics and replaced by the notion of curved space-time in the
Einsteinian theory.}. The hydrodynamic equations
(\ref{intro1})-(\ref{intro5}) or (\ref{intro6})-(\ref{intro10})
involving a barotropic equation of state, a long-range potential of
interaction and a friction force, have been introduced by Chavanis
\cite{gen,banach} at a general level. It was indicated that they 
could provide generalized models of chemotaxis and self-gravitating
Brownian particles.

The main difference between the chemotactic model
(\ref{intro1})-(\ref{intro5}) and the gravitational model
(\ref{intro6})-(\ref{intro10}) concerns the field equations
(\ref{intro3}) and (\ref{intro8}). In astrophysics, the gravitational
potential is determined instantaneously from the density of particles
through the Poisson equation (\ref{intro8}). In biology, the equation
(\ref{intro3}) determining the evolution of the chemical is more
complex and involves memory terms.  The chemical diffuses with a
diffusion coefficient $D_{c}$, is produced by the organisms at a rate
$h$ and is degraded at a rate $k$. Because of the term $\partial
c/\partial t$, the concentration of the chemical at time $t$ depends
on the concentration of the organisms at earlier times. In this paper,
we shall consider simplified models where the term $\partial
c/\partial t$ can be neglected.  This is valid in a limit of large
diffusivity of the chemical $D_{c}\rightarrow +\infty$
\cite{jl}. We first consider the case where there is no degradation
of the chemical ($k=0$). Then, assuming $h=\lambda D_{c}$ and taking
the limit $D_{c}\rightarrow +\infty$ with $\lambda=O(1)$, one gets
(see Appendix C of \cite{csbio})
\begin{equation}
\label{intro11}
\Delta c=-\lambda(\rho-\overline{\rho}),
\end{equation}
where $\overline{\rho}=(1/V)\int\rho d{\bf r}=M/V$ is the average
value of the density which is a conserved quantity. In that case, the
concentration of the chemical is given by a Poisson equation which
incorporates a sort of ``neutralizing background'' (played by
$\overline{\rho}$) like in the Jellium model of plasma physics 
\cite{alastuey}. Note that a similar term also arises in cosmology
when we take into account the expansion of the universe and work in a
comoving system of coordinates \cite{peebles}. We shall thus refer to
this model as the ``Newtonian model''. Then, we consider the case of a
finite degradation rate. Assuming $h=\lambda D_{c}$,
$k=k_{0}^{2}D_{c}$ and taking the limit $D_{c}\rightarrow +\infty$
with $\lambda=O(1)$ and $k_{0}=O(1)$, one gets (see Appendix C of \cite{csbio})
\begin{equation}
\label{intro12}
\Delta c-k_{0}^{2}c=-\lambda\rho.
\end{equation}
If we take formally $k_{0}=0$, we obtain a Poisson equation similar
to Eq. (\ref{intro8}) where $-c({\bf r},t)$ plays the role of
$\Phi({\bf r},t)$ and $\lambda$ plays the role of the gravitational
constant $S_{d}G$ (we recall that the geometrical factor $S_{d}$ is
the surface of a unit sphere in $d$ dimensions). However, Eq.
(\ref{intro12}) has been derived for $k_{0}\neq 0$ (for $k_{0}=0$ we
get Eq. (\ref{intro11})). This implies that the interaction is shielded
on a typical distance $k_{0}^{-1}$. This is similar to the Debye
shielding in plasma physics, to the Rossby shielding in geophysical
flows or to the Yukawa shielding in nuclear physics. We shall refer
to this model as the ``Yukawa model''.

In this paper, we perform a detailed linear dynamical stability
analysis of the chemotactic model
(\ref{intro1})-(\ref{intro3}). Specifically, we study the stability of
an infinite and homogeneous distribution of cells against chemotactic
collapse. This is similar to the classical Jeans stability analysis
for the barotropic Euler-Poisson system \cite{bt}. Indeed, the
``chemotactic collapse'' of biological populations is similar to the
``gravitational collapse'' in astrophysics (Jeans instability). There
are, however, two main differences with the classical Jeans
analysis. The first difference is the presence of a friction force
$-\xi {\bf u}$ in the Euler equation. As we shall see, this does not
change the onset of the instability but this affects the evolution of
the perturbation. The second difference arises from the different
nature of the field equations (\ref{intro3}) and (\ref{intro8}). We
recall that, in gravitational dynamics, an infinite and homogeneous
distribution of matter with $\rho=cst$ and ${\bf u}={\bf 0}$ is {\it
not} a stationary solution of the barotropic Euler-Poisson system
(\ref{intro6})-(\ref{intro8}) because we cannot satisfy simultaneously
the condition of hydrostatic equilibrium $\nabla
p(\rho)+\rho\nabla\Phi={\bf 0}$ reducing to $\nabla\Phi={\bf 0}$ and
the Poisson equation $\Delta\Phi=S_{d}G\rho\neq 0$. This leads to an
inconsistency in the mathematical analysis when studying the linear
dynamical stability of such a distribution: this is called the ``Jeans
swindle''
\cite{bt} \footnote{One possibility to avoid the Jeans swindle is to study the
linear dynamical stability of an {\it inhomogeneous} distribution of matter
in a finite domain (box) \cite{aaiso}. Alternatively, in cosmology,
the ``Jeans swindle'' is cured by the expansion of the universe
\cite{peebles}. Indeed, if we work in a comoving system of coordinates,
the usual Poisson equation $\Delta\Phi=4\pi G\rho$ is replaced by an
equation of the form $\Delta\phi=4\pi G a(t)^{2}\lbrack
\rho({\bf x},t)-\rho_{b}(t)\rbrack$ where the density $\rho({\bf
x},t)$ is replaced by the deviation $\rho({\bf x},t)-{\rho}_{b}(t)$ to
the mean density \cite{peebles}. Then, an infinite and homogeneous
distribution of matter with $\rho=\rho_{b}$ and $\phi=0$ is a steady
state of the equations of motion from which we can develop a rigorous
stability analysis. The expansion of the universe introduces a sort of
neutralizing background in the Poisson equation. Interestingly, the
same effect arises in the chemotactic model (\ref{intro11}) for a
completely different reason. Note finally that, in early models of
cosmology, some authors including Einstein himself have modified the
gravitational Poisson equation to the form
$\Delta\Phi-\lambda\Phi=4\pi G\rho$ by including a shielding term
\cite{pais}. This transformation was done in order to obtain a static
homogenous and isotropic universe. As we have seen, a similar
shielding effect arises naturally in the chemotactic model
(\ref{intro12}) due to the degradation of the chemical.}. By contrast,
{\it there is no ``Jeans swindle'' in the chemotactic problem!} 
Indeed, an infinite and homogeneous distribution of cells {\it is} a
steady state of the equations of motion (\ref{intro1})-(\ref{intro3})
corresponding to the condition $kc=h\rho$. For the ``Newtonian model''
(\ref{intro11}), this condition becomes $\rho=\overline{\rho}$ and for
the ``Yukawa model'' (\ref{intro12}), it becomes
$k_{0}^{2}c=\lambda\rho$.

In this paper, we study in detail the onset of the ``chemotactic
instability'' and its development in the linear regime. This study was
initiated in \cite{epjbjeans} at a general level, i.e. taking
into account the term $\partial c/\partial t$ in Eq. (\ref{intro3})
and allowing the coefficients in Eqs. (\ref{intro1})-(\ref{intro3}) to
depend on the concentration. However, this study focused on the
unstable modes and did not analyze in detail the evolution of the
stable modes. In the present paper, we make a complete study of both
stable and unstable modes but we restrict ourselves to the simplified
models (\ref{intro11}) and (\ref{intro12}). In the ``Newtonian model''
(\ref{intro11}), the only difference with the Jeans analysis is the
presence of the friction force $\xi$. In the ``Yukawa model''
(\ref{intro12}), the differences with the Jeans analysis are due to
the effects of the friction $\xi$ and of the shielding length
$k_{0}^{-1}$ generated by the degradation of the chemical.  We show
that the system is always stable for
\begin{equation}
\label{intro5ht}
c_{s}\ge (c_{s})_{crit}\equiv \left (\frac{\lambda\overline{\rho}}{k_{0}^{2}}\right )^{1/2},
\end{equation}
where $c_{s}\equiv (dp/d\rho)^{1/2}$ is the ``velocity of sound'' in
the medium (for specific equations of state, discussed in
Sec. \ref{sec_eos}, we can express the stability criterion
(\ref{intro5ht}) in terms of the density of cells).  Therefore, the
system is stable if the velocity of sound is above a certain threshold
fixed by the shielding length $k_{0}$. By contrast, for
$c_{s}<(c_{s})_{crit}$, the system is unstable for wavenumbers
\begin{equation}
\label{intro7ht}
k\le k_{m}\equiv \sqrt{k_{J}^{2}-k_{0}^{2}},
\end{equation}
where $k_{J}=(\lambda\overline{\rho}/c_{s}^{2})^{1/2}$ is the Jeans
wavenumber. In the Newtonian model, the condition $k_{0}=0$ implies
$(c_{s})_{crit}=+\infty$, so that the system is always unstable to
perturbations with sufficiently large wavelengths $k<k_{J}$. These
results are independent on $\xi$. The friction term only affects the
evolution of the perturbation. For $k<k_{m}(k_{0})$, the perturbation
grows exponentially rapidly, for $k_{m}(k_{0})<k<k_{c}(\xi,k_{0})$
(where $k_{c}$ is a friction-dependent wavenumber defined in the text)
it is damped exponentially rapidly without oscillating and for
$k>k_{c}(\xi,k_{0})$ it presents damped oscillations. More precisely,
we determine the growth rate of the unstable modes and the damping
rate, and oscillation frequency, of the stable modes as a function of
$\xi$ and $k_{0}^{-1}$. Owing to the above mentioned analogy between
chemotaxis and gravity, our stability analysis also applies to
self-gravitating Langevin particles
\cite{virial} provided that we make the ``Jeans swindle''.

\section{Jeans-type instability for a Newtonian potential}
\label{sec_jeans}

In this section, we study the linear dynamical stability of an infinite and
homogeneous stationary solution of the fluid equations
\begin{equation}
\label{jeans1}
\frac{\partial\rho}{\partial t}+\nabla\cdot (\rho {\bf u})=0,
\end{equation}
\begin{equation}
\label{jeans2} {\partial {\bf u}\over\partial t}+({\bf u}\cdot
\nabla){\bf u}=-{1\over\rho}\nabla p+\nabla c-\xi {\bf u},
\end{equation}
\begin{equation}
\label{jeans3}\Delta c=-\lambda (\rho-\overline{\rho}).
\end{equation}
We consider an infinite and homogeneous distribution of cells
$\rho=\overline{\rho}$, with no velocity ${\bf u}={\bf 0}$ and no
chemical $c=0$. This is an exact stationary solution of the fluid
equations (\ref{jeans1})-(\ref{jeans3}).  Linearizing
Eqs. (\ref{jeans1})-(\ref{jeans3}) around this steady state and
writing the perturbation in the form $\delta f({\bf r},t)\sim
e^{i({\bf k}\cdot {\bf r}-\omega t)}$, we readily obtain the
dispersion relation \cite{virial}:
\begin{equation}
\label{jeans4}
\omega (\omega+i\xi)=c_{s}^{2}k^{2}-\lambda\overline{\rho},
\end{equation}
where we have introduced the equivalent of the velocity of sound
$c_{s}^{2}=p'(\overline{\rho})$.  Setting $\sigma=-i\omega$, so that $\delta f\propto e^{\sigma t}$, the
dispersion relation can be rewritten
\begin{equation}
\label{jeans5}
\sigma^{2}+\xi\sigma+c_{s}^{2}k^{2}-\lambda\overline{\rho}=0.
\end{equation}
The solutions are $\sigma_{\pm}=(-\xi\pm \sqrt{\Delta})/2$
with $\Delta(k)=\xi^{2}-4(c_{s}^{2}k^{2}-\lambda\overline{\rho})$. If
$c_{s}^{2}k^{2}-\lambda\overline{\rho}<0$, then $\Delta>\xi^{2}>0$ and
the system is unstable since $\sigma_{+}=(-\xi+
\sqrt{\Delta})/2>0$. If $c_{s}^{2}k^{2}-\lambda\overline{\rho}>0$,
either (i) $\Delta<0$ implying $R_{e}(\sigma)=-\xi/2$ or (ii)
$0<\Delta<\xi^{2}$, implying $\sigma_{\pm}<0$, so the
system is stable. Therefore, the system is unstable if
\begin{equation}
\label{jeans6}
k<\left ({\lambda\overline{\rho}\over c_{s}^{2}}\right)^{1/2}\equiv k_{J},
\end{equation}
and stable otherwise. The critical value $k_{J}$ is similar to the
Jeans wavenumber in astrophysics. We note that the threshold of
instability does not depend on the friction parameter $\xi$. Note also
that for negative chemotaxis (chemorepulsion) obtained by replacing
$+\nabla c$ by $-\nabla c$ in Eq. (\ref{jeans2}), an infinite and
homogeneous distribution of particles is always stable.

If we consider the case $\xi=0$ (Euler), the fluid equations
(\ref{jeans1})-(\ref{jeans3}) are similar to the Euler-Poisson system
and the dispersion relation becomes
\begin{equation}
\label{jeans7}
\omega^{2}=c_{s}^{2}k^{2}-\lambda\overline{\rho}.
\end{equation}
For $k>k_{J}$, the perturbation undergoes undamped oscillations with
pulsation $\omega=c_{s}(k^{2}-k_{J}^{2})^{1/2}$. For $k<k_{J}$, the
perturbation increases exponentially rapidly with a growth rate
$\gamma=c_{s}(k_{J}^{2}-k^{2})^{1/2}$. For $c_{s}=0$,
$k_{J}\rightarrow +\infty$ and the system is unstable for all
wavenumbers. The growth rate of the perturbation is
$\gamma=\sqrt{\lambda\overline{\rho}}$ independent on $k$. For
$c_{s}\rightarrow +\infty$, $k_{J}\rightarrow 0$ and the system is
stable for all wavenumbers. The pulsation is $\omega=c_{s}k$. If we
now consider the case $\xi\rightarrow +\infty$ (Smoluchowski), the
fluid equations (\ref{jeans1})-(\ref{jeans3}) reduce to the
generalized Smoluchowski-Poisson system
\begin{equation}
\label{jeans8}
\frac{\partial\rho}{\partial t}=\nabla\cdot \left\lbrack \frac{1}{\xi}(\nabla p-\rho\nabla c)\right\rbrack,
\end{equation}
\begin{equation}
\label{jeans9}\Delta c=-\lambda (\rho-\overline{\rho}),
\end{equation}
and the dispersion relation becomes
\begin{equation}
\label{jeans10}
i\xi\omega=c_{s}^{2}k^{2}-\lambda\overline{\rho}.
\end{equation}
For $k>k_{J}$, the perturbation decays exponentially rapidly with a
damping rate $\gamma=-{c_{s}^{2}}(k^{2}-k_{J}^{2})/\xi$. For
$k<k_{J}$, the perturbation increases exponentially rapidly with a
growth rate $\gamma={c_{s}^{2}}(k_{J}^{2}-k^{2})/\xi$. For $c_{s}=0$, $k_{J}\rightarrow +\infty$ and the system is unstable for all wavenumbers. The growth rate of the perturbation is $\gamma=\lambda\overline{\rho}/\xi$ independent on $k$. For $c_{s}\rightarrow +\infty$, $k_{J}\rightarrow 0$ and the system is stable for all wavenumbers. The damping rate is $\gamma=-c_{s}^{2}k^{2}/\xi$.

\begin{figure}
\centerline{
\psfig{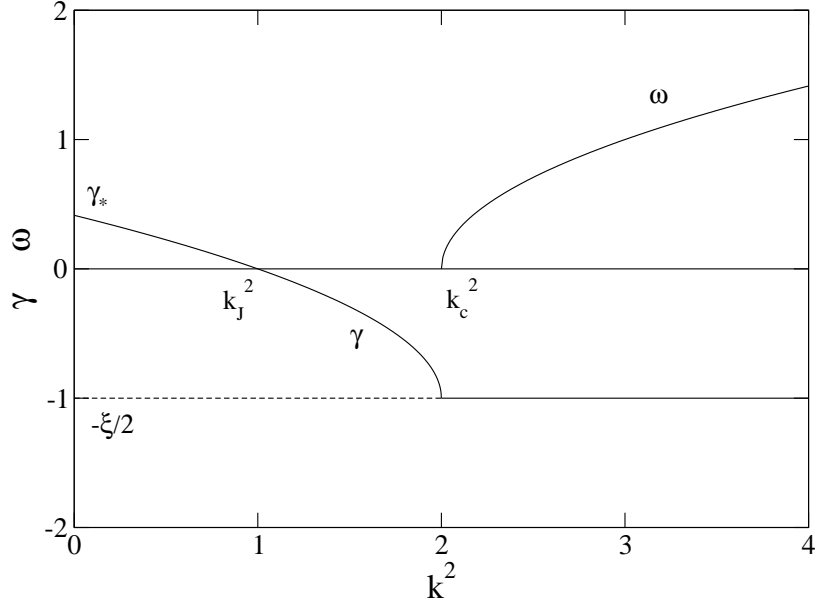}}
\caption{In summary, a homogeneous distribution is unstable for $k<k_{J}$ and stable for
$k>k_{J}$. For $k<k_{J}$, the perturbation grows exponentially rapidly.
For $k_{J}<k<k_{c}$, the perturbation is damped exponentially rapidly without oscillating. For $k>k_{c}$, the perturbation undergoes damped oscillations. We have taken $F=1$, $k_{J}=1$ and $\xi=2$.} \label{jeans}
\end{figure}

Let us now consider the general case of an arbitrary friction. There
are two relevant wavenumbers in the problem: the Jeans wavenumber
(\ref{jeans6}) and the wavenumber
\begin{equation}
\label{jeans11}
k_{c}\equiv \left (k_{J}^{2}+k_{d}^{2}\right )^{1/2},
\end{equation}
where
\begin{equation}
\label{jeans12}
k_{d}\equiv {\xi\over 2c_{s}},
\end{equation}
is a wavenumber constructed with the friction coefficient and the
velocity of sound. We can define a dimensionless number
\begin{equation}
\label{jeans13}
F=\left (\frac{k_{d}}{k_{J}}\right )^{2}=\frac{\xi^{2}}{4\lambda\overline{\rho}},
\end{equation}
which measures the strength of the friction force (a similar parameter
was introduced in \cite{virial} for inhomogeneous distributions). It is
independent on the equation of state $p(\rho)$ and it can be written
$F\sim (\xi t_{D})^{2}$ where $t_{D}\sim 1/\sqrt{\overline{\rho}
\lambda}$ is a typical dynamical time. Thus, $\sqrt{F}$ is the ratio of the
dynamical time on the friction time $\tau\sim 1/\xi$. In terms of this
parameter, the wavenumber (\ref{jeans11}) can be written
$k_{c}=k_{J}\sqrt{1+F}$. The behaviour of the perturbation can be
analyzed in terms of these wavenumbers: (i) If $\Delta<0$, the
perturbation undergoes damped oscillations with pulsation
$\omega=c_{s} (k^{2}-k_{c}^{2})^{1/2}$ and decay rate
$\gamma=-\xi/2$. This stable regime corresponds to wavenumbers
$k>k_{c}$. (ii) If $0<\Delta<\xi^{2}$, the perturbation decays
exponentially rapidly with a damping rate $\gamma=-\xi/2+c_{s}(
k_{c}^{2} -k^{2})^{1/2}$ without oscillating. This stable regime
corresponds to wavenumbers $k_{J}<k<k_{c}$. For $k=k_{J}$, we have
$\gamma=0$ and for $k=k_{c}$, we have $\gamma=-\xi/2$. (iii) If
$\Delta>\xi^{2}$, the perturbation increases exponentially rapidly
with a growth rate $\gamma=-\xi/2+c_{s}(k_{c}^{2}-k^{2})^{1/2}$. This
unstable regime corresponds to wavenumbers $k<k_{J}$. The growth rate
is maximum for $k_{*}=0$ and its value is
$\gamma_{*}=-\xi/2+c_{s}k_{c}$. These results are summarized in
Fig. \ref{jeans}.

It is interesting to determine how the results depend on the friction parameter and on the velocity of sound. To simplify the notations, we define $T\equiv c_{s}^{2}$. Then, we obtain
\begin{equation}
\label{jeans14}
k_{J}(T)=\left ({\lambda\overline{\rho}\over T}\right )^{1/2},
\end{equation}
\begin{equation}
\label{jeans15}
k_{c}(T,\xi)=k_{J}(T)\sqrt{1+F},
\end{equation}
\begin{equation}
\label{jeans16}
k_{*}=0, \qquad {2\over\xi}\gamma_{*}(\xi)=-1+\sqrt{1+{1\over F}}
\end{equation}
\begin{equation}
\label{jeans17}
{2\over\xi}\gamma(k,T,\xi)=-1+\sqrt{1+{1\over F}}\left\lbrack 1-\left ({k\over k_{c}}\right )^{2}\right\rbrack^{1/2}, \qquad (k<k_{c}).
\end{equation}
\begin{equation}
\label{jeans18}
{2\over\xi}\omega(k,T,\xi)=\sqrt{1+{1\over F}}\left\lbrack \left ({k\over k_{c}}\right )^{2}-1\right\rbrack^{1/2}, \qquad \gamma=-\xi/2, \qquad (k>k_{c}).
\end{equation}

For $T=0$, we find that $k_{J}\rightarrow +\infty$ so that the system is unstable for all wavenumbers. The growth rate of the perturbation is
\begin{equation}
\label{jeans19}
{2\gamma\over\xi}=-1+\sqrt{1+{1\over F}},
\end{equation}
independent on $k$.  For $T\rightarrow +\infty$, if $\xi$ is finite,
we get $k_{J}=k_{c}=0$ so that the system is stable for all wavenumbers. The
pulsation is $\omega=\sqrt{T}k$ and the damping rate $\gamma=-\xi/2$.

\begin{figure}
\centerline{
\psfig{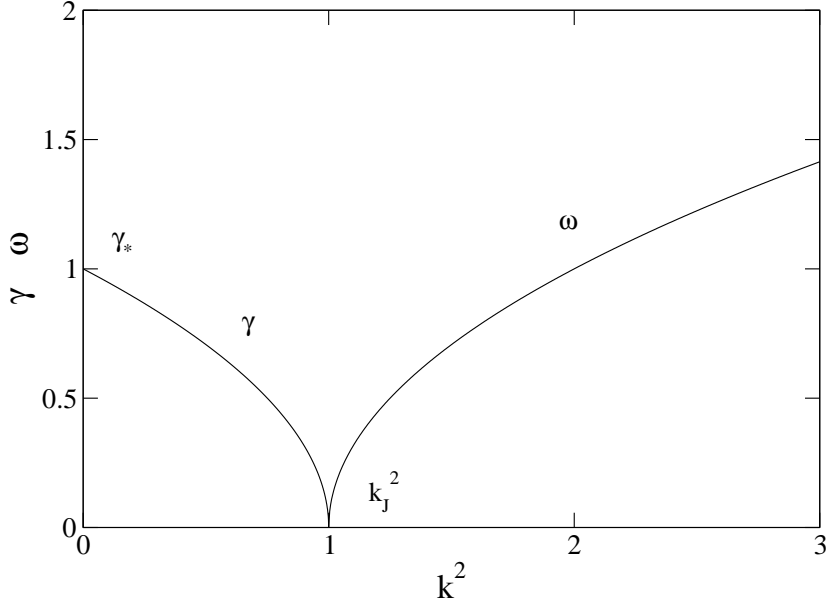}}
\caption{The limit $\xi=0$.  We have taken $k_{J}=1$ and $\lambda\overline{\rho}=1$.} \label{xi0jeans}
\end{figure}

For $\xi=0$, we find that $k_{c}=k_{J}$ and
\begin{equation}
\label{jeans20}
\gamma(k,T)=\sqrt{\lambda\overline{\rho}}\left\lbrack 1-\left ({k\over k_{J}}\right )^{2}\right\rbrack^{1/2}, \qquad (k<k_{J})
\end{equation}
\begin{equation}
\label{jeans21}
\omega(k,T)=\sqrt{\lambda\overline{\rho}}\left\lbrack \left ({k\over k_{J}}\right )^{2}-1\right\rbrack^{1/2}, \qquad \gamma=0, \qquad (k>k_{J}).
\end{equation}
These results are summarized in
Fig. \ref{xi0jeans}.
For $T=0$, we find that $k_{J}\rightarrow +\infty$ so that the system
is unstable for all wavenumbers. The growth rate of the perturbation
is $\gamma=\sqrt{\lambda\overline{\rho}}$. For $T\rightarrow +\infty$,
we get $k_{J}=0$ so that the system is stable for all wavenumbers. The
pulsation is $\omega=\sqrt{T}k$.

\begin{figure}
\centerline{
\psfig{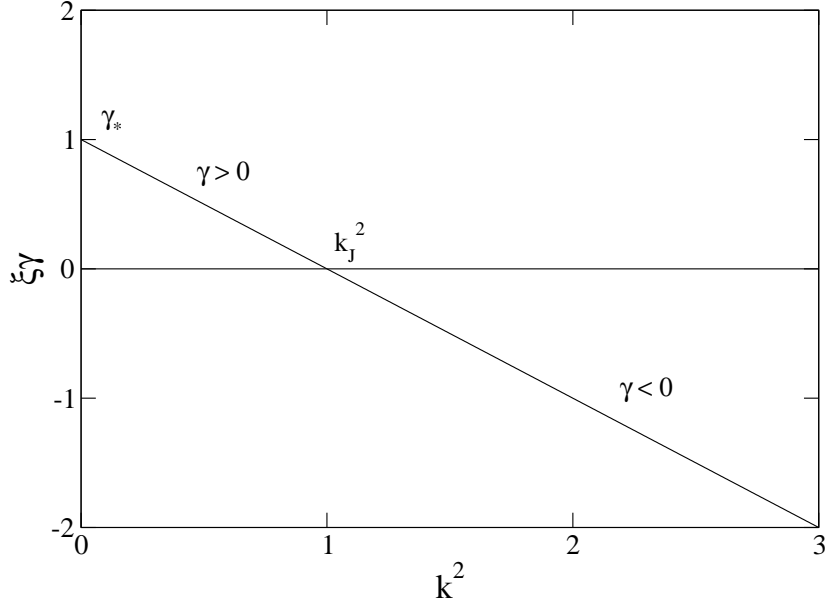}}
\caption{The limit $\xi\rightarrow +\infty$. We have taken $k_{J}=1$ and $\lambda\overline{\rho}=1$} \label{xiGjeans}
\end{figure}

For $\xi\rightarrow +\infty$, we find that $k_{c}\rightarrow +\infty$ and
\begin{equation}
\label{jeans22}
\gamma(k,T)={\lambda\overline{\rho}\over\xi}\left (1-{k^{2}\over k_{J}^{2}}\right ).
\end{equation}
These results are summarized in
Fig. \ref{xiGjeans}. For $T=0$, we find that $k_{J}\rightarrow +\infty$ so that the system
is unstable for all wavenumbers. The growth rate of the perturbation
is $\gamma={\lambda\overline{\rho}/\xi}$. For $T\rightarrow +\infty$,
we get $k_{J}=0$ so that the system is stable for all wavenumbers. The
damping rate is $\gamma=-{T}k^{2}/\xi$.

If we now consider the stability problem of
Eqs. (\ref{jeans1})-(\ref{jeans3}) in a two-dimensional periodic
domain of size $L$, the wavenumbers can be written ${\bf k}={2\pi\over
L}(m,n)$ where $m$, $n$ are positive integers with $(m,n)\neq
(0,0)$. In that case, the condition of instability (\ref{jeans6})
becomes
\begin{equation}
\label{jeans23}
m^{2}+n^{2}<{\lambda\overline{\rho}L^{2}\over 4\pi^{2}c_{s}^{2}}.
\end{equation}
A necessary condition of instability is therefore
$\lambda\overline{\rho}L^{2}/(4\pi^{2}c_{s}^{2})>1$.  For an equation
of state of the form $p(\rho)=\rho T$, where $T$ plays the role of a
temperature, the velocity of sound is $c_{s}^{2}=T$ and the necessary
condition of instability can be written
\begin{equation}
\label{jeans25}
T<T_{c}={\lambda\overline{\rho}L^{2}\over 4\pi^{2}},
\end{equation}
where $T_{c}$ is a critical temperature. For $T>T_{c}$, there is no
chemotactic collapse: the ``gas'' of cells remains spatially uniform
and diffuse. For $T<T_{c}$, the distribution of cells is unstable and
the number of unstable modes increases as $T$ decreases yielding more
and more clusters. This instability has been illustrated numerically
in \cite{csbio} by solving the $N$-body equations of motion in a
two-dimensional periodic domain.

\section{Jeans-type instability criterion for a Yukawa potential}
\label{sec_yukawa}

We now consider the linear dynamical stability of an infinite and
homogeneous stationary solution of the fluid equations
\begin{equation}
\label{yukawa1}
\frac{\partial\rho}{\partial t}+\nabla\cdot (\rho {\bf u})=0,
\end{equation}
\begin{equation}
\label{yukawa2} {\partial {\bf u}\over\partial t}+({\bf u}\cdot
\nabla){\bf u}=-{1\over\rho}\nabla p+\nabla c-\xi {\bf u},
\end{equation}
\begin{equation}
\label{yukawa3}\Delta c-k_{0}^{2}c=-\lambda \rho.
\end{equation}
Comparing with Eq. (\ref{jeans3}), we see that the Laplacian is
replaced by the operator $\Delta-k_{0}^{2}$. This implies that the
interaction, mediated by the concentration $c$ of the chemical, is
screened on a distance $k_{0}^{-1}$ where $k_{0}=(k/D_{c})^{1/2}$ [$k$
should not be confused here with the wavenumber].  A uniform
distribution of cells and secreted chemicals whose concentrations
satisfy the relation $k_{0}^{2}c=\lambda \overline{\rho}$ is an
exact stationary solution of Eqs.  (\ref{yukawa1})-(\ref{yukawa3}).
Considering a small perturbation around this steady state, we find
that the dispersion relation replacing Eq. (\ref{jeans5}) is \cite{virial}:
\begin{equation}
\label{yukawa4}
\sigma^{2}+\xi\sigma+k^{2}\left (c_{s}^{2}-{\lambda\overline{\rho}\over k^{2}+k_{0}^{2}}\right )=0.
\end{equation}
The solutions are $\sigma_{\pm}=(-\xi\pm \sqrt{\Delta})/2$
with
\begin{equation}
\label{yukawa5}
\Delta(k)=\xi^{2}-4k^{2}\left (c_{s}^{2}-{\lambda\overline{\rho}\over k^{2}+k_{0}^{2}}\right ).
\end{equation}
Repeating the arguments following Eq. (\ref{jeans5}), we find that the system is unstable if
\begin{equation}
\label{yukawa6}
c_{s}^{2}<{\lambda\overline{\rho}\over k^{2}+k_{0}^{2}},
\end{equation}
and stable otherwise. A necessary condition for instability is that
\begin{equation}
\label{yukawa7}
c_{s}^{2}<(c_{s}^{2})_{crit}\equiv {\lambda\overline{\rho}\over k_{0}^{2}}.
\end{equation}
When this condition is fulfilled, the range of unstable wavelengths is
\begin{equation}
\label{yukawa8}
k\le k_{m}\equiv \sqrt{k_{J}^{2}-k_{0}^{2}}.
\end{equation}
We see that for $k_{0}\neq 0$, the instability is shifted to larger
wavelengths than the Jeans length.

If we consider the case $\xi=0$ (Euler),  the dispersion relation becomes
\begin{equation}
\label{yukawa9}
\sigma^{2}=c_{s}^{2} k^{2}{k_{m}^{2}-k^{2}\over k_{0}^{2}+k^{2}}.
\end{equation}
For $k>k_{m}$, the perturbation undergoes undamped oscillations with
pulsation $\omega=\sqrt{-\sigma^{2}}$. For $k<k_{m}$, the
perturbation increases exponentially rapidly with a growth rate
$\gamma=\sqrt{\sigma^{2}}$. If we consider the case
$\xi\rightarrow +\infty$ (Smoluchowski), the dispersion relation becomes
\begin{equation}
\label{yukawa10}
\xi\sigma=c_{s}^{2} k^{2}{k_{m}^{2}-k^{2}\over k_{0}^{2}+k^{2}}.
\end{equation}
For $k>k_{m}$, the perturbation decays exponentially rapidly with a
damping rate $\gamma=\sigma<0$. For
$k<k_{m}$, the perturbation increases exponentially rapidly with a
growth rate $\gamma=\sigma>0$.

\begin{figure}
\centerline{
\psfig{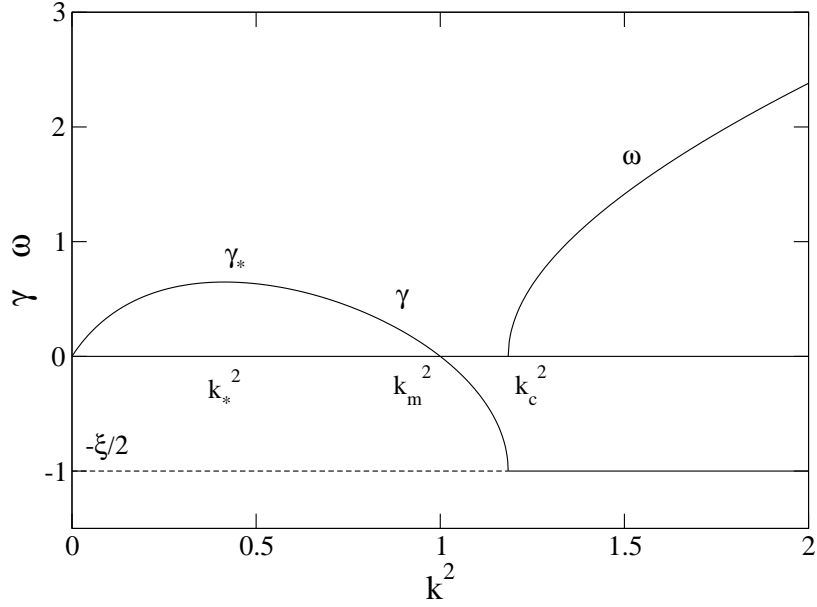}}
\caption{Growth rate and pulsation as a function of the wavenumber.  We have taken $T/T_{c}=1/2$, $F=0.05$, $k_{0}=1$ and $\xi=2$} \label{yukawa}
\end{figure}

Let us now consider the general case of an arbitrary friction. We
introduce a critical wavenumber $k_{c}$ defined by
\begin{equation}
\label{yukawa11}
2k_{c}^{2}=k_{m}^{2}+k_{d}^{2}+\sqrt{(k_{m}^{2}+k_{d}^{2})^{2}+4k_{0}^{2}k_{d}^{2}}.
\end{equation}
This expression generalizes Eq. (\ref{jeans11}) to the case $k_{0}\neq 0$. We also introduce the wavenumber $k_{s}$ defined by
\begin{equation}
\label{yukawa12}
2k_{s}^{2}=-(k_{m}^{2}+k_{d}^{2})+\sqrt{(k_{m}^{2}+k_{d}^{2})^{2}+4k_{0}^{2}k_{d}^{2}}.
\end{equation}
The behaviour of the perturbation can be analyzed in terms of these
wavenumbers: (i) If $\Delta<0$, the perturbation undergoes damped
oscillations with pulsation
\begin{equation}
\label{yukawa13}
\omega={\xi\over 2}\sqrt{{k^{2}(k^{2}-k_{m}^{2})\over k_{d}^{2}(k^{2}+k_{0}^{2})}-1},
\end{equation}
or equivalently
\begin{equation}
\label{yukawa14}
\omega=c_{s}\left\lbrack \frac{(k^{2}-k_{c}^{2})(k^{2}+k_{s}^{2})}{k^{2}+k_{0}^{2}}\right \rbrack^{1/2},
\end{equation}
and decay rate $\gamma=-\xi/2$. This stable regime corresponds to
wavenumbers $k>k_{c}$. (ii) If $0<\Delta<\xi^{2}$, the perturbation
decays exponentially rapidly with a damping rate
\begin{equation}
\label{yukawa15}
\gamma=-{\xi\over 2}+{\xi\over 2}\sqrt{1-{k^{2}(k^{2}-k_{m}^{2})\over k_{d}^{2}(k^{2}+k_{0}^{2})}},
\end{equation}
or equivalently
\begin{equation}
\label{yukawa16}
\gamma=-{\xi\over 2}+c_{s}\left\lbrack {(k_{c}^{2}-k^{2})(k_{s}^{2}+k^{2})\over k^{2}+k_{0}^{2}}\right\rbrack^{1/2},
\end{equation}
without oscillating. This stable regime corresponds to wavenumbers
$k_{m}<k<k_{c}$. For $k=k_{c}$, we have $\gamma=-\xi/2$ and for
$k=k_{m}$ we have $\gamma=0$. (iii) If $\Delta>\xi^{2}$, the
perturbation increases exponentially rapidly with a growth rate
$\gamma$ given by Eq. (\ref{yukawa15}). This unstable regime
corresponds to wavenumbers $k<k_{m}$. The growth rate is maximum for
\begin{equation}
\label{yukawa17}
k_{*}=\sqrt{k_{0}(k_{J}-k_{0})},
\end{equation}
and its value is
\begin{equation}
\label{yukawa18}
\gamma_{*}=-{\xi\over 2}+c_{s}\sqrt{k_{d}^{2}+(k_{J}-k_{0})^{2}}.
\end{equation}
In summary, a homogeneous distribution is unstable for $k<k_{m}$ and stable for
$k>k_{m}$. For $k<k_{m}$, the perturbation grows exponentially rapidly.
For $k_{m}<k<k_{c}$, the perturbation is damped exponentially rapidly without oscillating. For $k>k_{c}$, the perturbation undergoes damped oscillations.
These results are summarized in
Fig. \ref{yukawa}.

\begin{figure}
\centerline{
\psfig{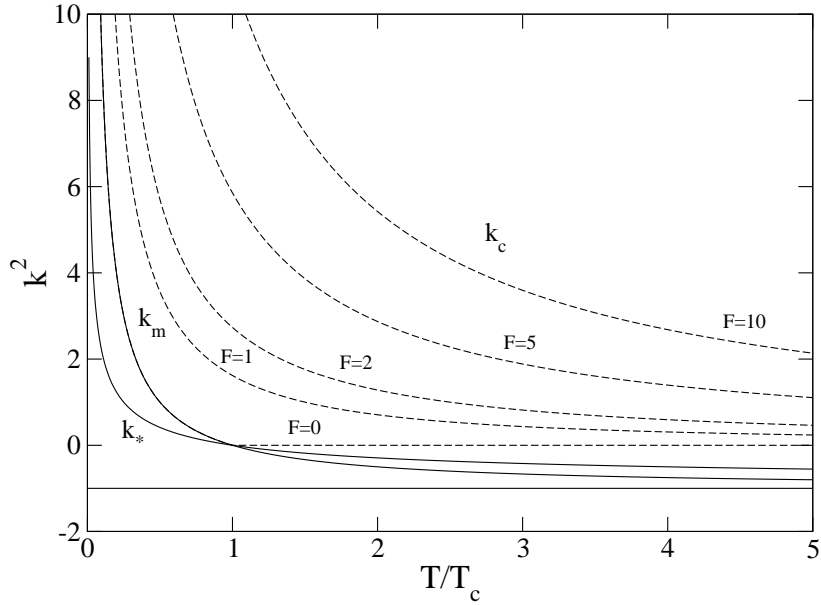}}
\caption{Dependence of the
characteristic scales with the temperature. The growth rate is maximum for $k=k_{*}(T)$ and the system is stable for $k>k_{m}(T)$. Oscillations appear for $k>k_{c}(T,F)$. For $T\rightarrow 0$, we have $k_{c}^{2}/k_{0}^{2}\sim (1+F)T_{c}/T$ and for $T\rightarrow +\infty$, we have  $k_{c}^{2}/k_{0}^{2}\sim F T_{c}/T$. Note that for $F=0$, $k_{c}=k_{m}$ for $T<T_{c}$ and $k_{c}=0$ for $T>T_{c}$.  } \label{kT}
\end{figure}

\begin{figure}
\centerline{
\psfig{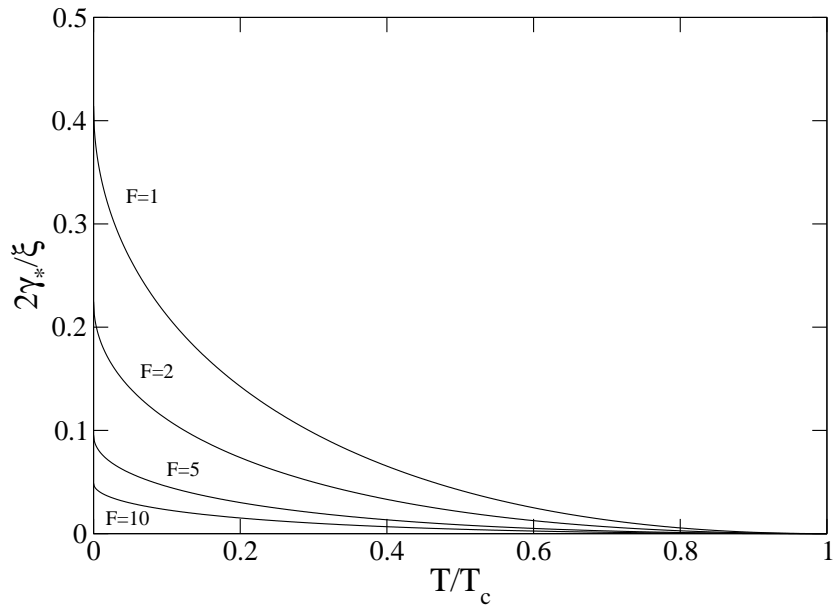}}
\caption{Maximum growth rate as a function of the temperature for different values of the friction parameter $F$. } \label{gammastar}
\end{figure}

It is interesting to determine how the results depend on the friction
parameter and on the velocity of sound (see Figs. \ref{kT} and
\ref{gammastar}). To simplify the notations, we define $T\equiv
c_{s}^{2}$. Then, we obtain
\begin{equation}
\label{yukawa19}
k_{J}(T)/k_{0}=\left ({T_{c}\over T}\right )^{1/2},\quad k_{m}(T)/k_{0}=\left ({T_{c}\over T}-1\right )^{1/2},\quad k_{d}(\xi,T)/k_{0}=\sqrt{F}\left ({T_{c}\over T}\right )^{1/2},
\end{equation}
\begin{equation}
\label{yukawa22}
2k_{c}^{2}(\xi,T)/k_{0}^{2}=(1+F){T_{c}\over T}-1+\sqrt{\left\lbrack (1+F){T_{c}\over T}-1\right\rbrack^{2}+4F{T_{c}\over T}},
\end{equation}
\begin{equation}
\label{yukawa23}
2k_{s}^{2}(\xi,T)/k_{0}^{2}=1-(1+F){T_{c}\over T}+\sqrt{\left\lbrack (1+F){T_{c}\over T}-1\right\rbrack^{2}+4F{T_{c}\over T}},
\end{equation}
\begin{equation}
\label{yukawa24}
k_{*}(T)/k_{0}=\left\lbrack \left ({T_{c}\over T}\right )^{1/2}-1\right\rbrack^{1/2},
\end{equation}
\begin{equation}
\label{yukawa25}
{2\over \xi}\gamma_{*}(\xi,T)=-1+\sqrt{1+{1\over F}\left\lbrack 1-\left ({T\over T_{c}}\right )^{1/2}\right\rbrack^{2}},
\end{equation}
\begin{equation}
\label{yukawa26}
{2\over \xi}\omega(k,\xi,T)=\sqrt{{k^{2}\lbrack k^{2}-k_{0}^{2}(T_{c}/T-1)\rbrack\over k_{0}^{2}F(T_{c}/T)(k^{2}+k_{0}^{2})}-1}, \quad \gamma=-{\xi\over 2}, \quad (k>k_{c}),
\end{equation}
\begin{equation}
\label{yukawa27}
{2\over \xi}\gamma(k,\xi,T)=-1+\sqrt{1-{k^{2}\lbrack k^{2}-k_{0}^{2}(T_{c}/T-1)\rbrack\over k_{0}^{2}F(T_{c}/T)(k^{2}+k_{0}^{2})}},  \quad (k<k_{c}).
\end{equation}

For $T=T_{c}$, we have
\begin{equation}
\label{yukawa28}
k_{J}=k_{0},\qquad k_{m}=0,\qquad k_{d}=k_{0}\sqrt{F},
\end{equation}
\begin{equation}
\label{yukawa29}
2k_{c}^{2}/k_{0}^{2}=F+\sqrt{F^{2}+4F}, \qquad 2k_{s}^{2}/k_{0}^{2}=-F+\sqrt{F^{2}+4F},
\end{equation}
\begin{equation}
\label{yukawa30}
{2\over \xi}\omega=\sqrt{{k^{4}\over k_{0}^{2}F(k^{2}+k_{0}^{2})}-1}, \quad \gamma=-{\xi\over 2}, \quad (k>k_{c}),
\end{equation}
\begin{equation}
\label{yukawa31}
{2\over \xi}\gamma=-1+\sqrt{1-{k^{4}\over k_{0}^{2}F(k^{2}+k_{0}^{2})}},  \quad (k<k_{c}).
\end{equation}
The system is stable for all wavelengths. The wavenumber marking the appearance of  oscillations is $k_{c}(F)$. It behaves like $k_{c}/k_{0}\sim F^{1/4}$ for $F\rightarrow 0$ and $k_{c}/k_{0}\sim F^{1/2}$ for $F\rightarrow +\infty$.

For $T=0$, we find that $k_{m}\rightarrow +\infty$ so that the system is always unstable. The growth rate of the perturbation is given by
\begin{equation}
\label{yukawa32}
{2\gamma\over\xi}=-1+\sqrt{1+{k^{2}\over F(k_{0}^{2}+k^{2})}}.
\end{equation}
It increases and tends asymptotically to its maximum value
\begin{equation}
\label{yukawa33}
{2\gamma_{*}\over\xi}=-1+\sqrt{1+{1\over F}}.
\end{equation}
For $T>T_{c}$, the system is always stable ($k_{m}^{2}<0$). For
$k>k_{c}$, the perturbation undergoes damped oscillations with
pulsation (\ref{yukawa13}) and decay rate $\gamma=-{\xi}/{2}$. For
$k<k_{c}$, the perturbation decays exponentially with rate
(\ref{yukawa15}) tending to $\gamma=0$ for $k=0$. 
For $T\rightarrow +\infty$ and $F$ finite, 
\begin{equation}
\label{yukawa34}
k_{J}=0,\qquad k_{m}^{2}/k_{0}^{2}=-1,\qquad k_{d}=0, \qquad k_{c}=0, \qquad k_{s}=k_{0}.
\end{equation}
The perturbation oscillates with a pulsation $\omega=\sqrt{T}k$,
and is damped with a rate $\gamma=-\xi/2$. The case $F\rightarrow
+\infty$ is treated in Sec. \ref{sec_yukawaxiinf}.

\subsection{The case $\xi=0$}
\label{sec_yukawaxi0}

\begin{figure}
\centerline{
\psfig{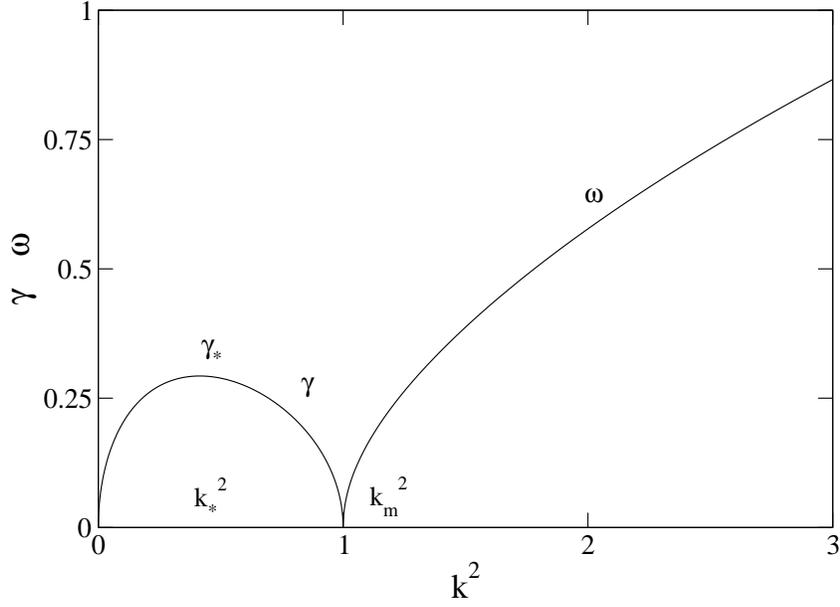}}
\caption{Growth rate and pulsation as a function of the wavenumber for $\xi=0$. We have taken $T/T_{c}=1/2$, $\lambda\overline{\rho}=1$, $k_{0}=1$.} \label{xi0yukawa}
\end{figure}

For $\xi=0$ (Euler) and $T\le T_{c}$, we have
\begin{equation}
\label{yukawa37}
k_{m}(T)=k_{c}(T)=k_{0}\left ({T_{c}\over T}-1\right )^{1/2}, \qquad k_{d}=k_{s}=0.
\end{equation}
For $k<k_{m}$, the system is unstable and the growth rate is
\begin{equation}
\label{yukawa39}
\gamma=\sqrt{T}\left\lbrack {k^{2}(k_{m}^{2}-k^{2})\over k^{2}+k_{0}^{2}}\right\rbrack^{1/2}.
\end{equation}
It is maximum for
\begin{equation}
\label{yukawa40}
k_{*}(T)/k_{0}=\left\lbrack \left ({T_{c}\over T}\right )^{1/2}-1\right\rbrack^{1/2},
\end{equation}
with value
\begin{equation}
\label{yukawa41}
\gamma_{*}(T)=k_{0}\sqrt{T_{c}}\left\lbrack 1-\left ({T\over T_{c}}\right )^{1/2}\right\rbrack.
\end{equation}
For $k>k_{m}$, the system is stable and the perturbation undergoes oscillations with pulsation
\begin{equation}
\label{yukawa41b}
\omega=\sqrt{T}\left\lbrack {k^{2}(k^{2}-k_{m}^{2})\over k^{2}+k_{0}^{2}}\right\rbrack^{1/2}.
\end{equation}
These results are summarized in Fig. \ref{xi0yukawa}.
For $T=T_{c}$, $k_{m}=0$. The system is always stable and the pulsation is
\begin{equation}
\label{yukawa42}
\omega=\sqrt{T_{c}}{k^{2}\over \sqrt{k^{2}+k_{0}^{2}}}.
\end{equation}
For $T=0$, $k_{m}\sim k_{0}(T_{c}/T)^{1/2}\rightarrow +\infty$. The system is always unstable and the growth rate is
\begin{equation}
\label{yukawa43}
\gamma=\sqrt{T_{c}}{k_{0}k\over \sqrt{k^{2}+k_{0}^{2}}}.
\end{equation}
It increases and tends, for $k\rightarrow +\infty$, to its maximum
value $\gamma_{*}=\sqrt{T_{c}}k_{0}=\sqrt{\lambda\overline{\rho}}$.
For $\xi=0$ and $T>T_{c}$, we have
\begin{equation}
\label{yukawa45}
k_{s}^{2}=-k_{m}^{2}=k_{0}^{2}\left (1-{T_{c}\over T}\right ), \qquad k_{d}=k_{c}=0.
\end{equation}
The system is always stable and the pulsation is
\begin{equation}
\label{yukawa47}
\omega=\sqrt{T}\left\lbrack {k^{2}(k^{2}+k_{s}^{2})\over k^{2}+k_{0}^{2}}\right\rbrack^{1/2}.
\end{equation}
For $T\rightarrow +\infty$, $k_{s}\rightarrow k_{0}$ and
$\omega\sim\sqrt{T}k$.

\subsection{The case $\xi\rightarrow +\infty$}
\label{sec_yukawaxiinf}

\begin{figure}
\centerline{
\psfig{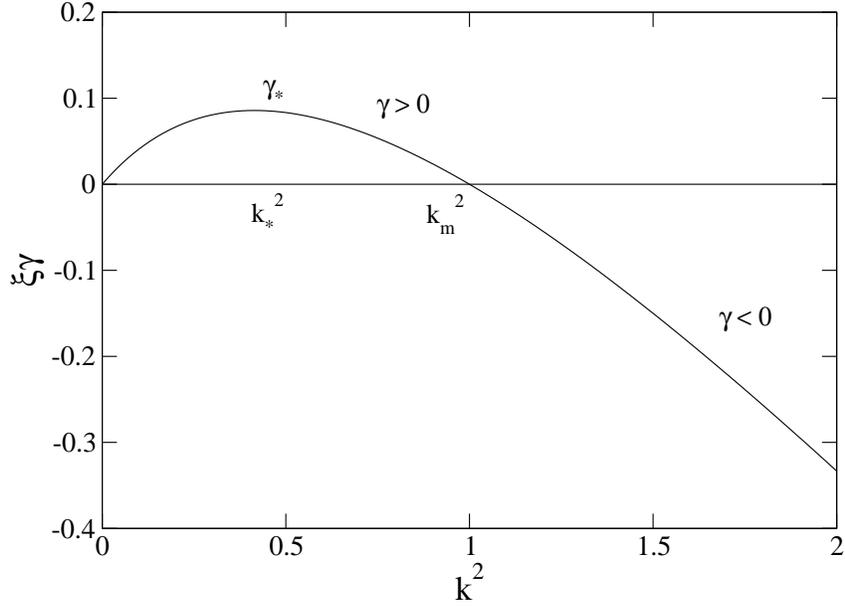}}
\caption{Growth rate as a function of the wavenumber for $\xi\rightarrow +\infty$. We have taken $T/T_{c}=1/2$, $\lambda\overline{\rho}=1$, $k_{0}=1$.} \label{xiGyukawa}
\end{figure}

For $\xi\rightarrow +\infty$ (Smoluchowski) and $T\le T_{c}$, we have
\begin{equation}
\label{yukawa49}
k_{m}(T)=k_{0}\left ({T_{c}\over T}-1\right )^{1/2},\qquad k_{c}\sim k_{d}\rightarrow +\infty, \qquad k_{s}=0.
\end{equation}
The rate of the exponential evolution is
\begin{equation}
\label{yukawa51}
\gamma={T\over \xi}{k^{2}(k_{m}^{2}-k^{2})\over k^{2}+k_{0}^{2}}.
\end{equation}
For $k>k_{m}$, the perturbation is damped (stable) and the decay rate behaves like $\gamma\sim -(T/\xi)k^{2}$ for $k\rightarrow +\infty$. For $k<k_{m}$ the perturbation increases (unstable). The growth rate is maximum for
\begin{equation}
\label{yukawa52}
k_{*}(T)/k_{0}=\left\lbrack \left ({T_{c}\over T}\right )^{1/2}-1\right\rbrack^{1/2},
\end{equation}
with value
\begin{equation}
\label{yukawa53}
\gamma_{*}(T)={k_{0}^{2}T_{c}\over \xi}\left\lbrack 1-\left ({T\over T_{c}}\right )^{1/2}\right\rbrack^{2}.
\end{equation}
These results are summarized in Fig. \ref{xiGyukawa}.

For $T=T_{c}$, $k_{m}=0$. The system is always stable and the perturbation decreases with a decay rate
\begin{equation}
\label{yukawa54}
\gamma=-{T_{c}\over \xi}{k^{4}\over k^{2}+k_{0}^{2}}.
\end{equation}
For $T=0$, $k_{m}\rightarrow +\infty$. The system is always unstable and the growth rate is
\begin{equation}
\label{yukawa55}
\gamma={k_{0}^{2}T_{c}\over\xi}{k^{2}\over {k^{2}+k_{0}^{2}}}.
\end{equation}
It increases and tends, for $k\rightarrow +\infty$, to its maximum
value $\gamma_{*}={k_{0}^{2}T_{c}/\xi}={\lambda\overline{\rho}/\xi}$.
For $\xi\rightarrow +\infty$ and $T>T_{c}$, we have
\begin{equation}
\label{yukawa57}
k_{m}^{2}=k_{0}^{2}\left ({T_{c}\over T}-1\right )<0, \qquad k_{c}\sim k_{d}\rightarrow +\infty, \qquad k_{s}=0.
\end{equation}
The system is always stable and the decay rate is
\begin{equation}
\label{yukawa59}
\gamma={T\over \xi}{k^{2}(k_{m}^{2}-k^{2})\over k^{2}+k_{0}^{2}}<0.
\end{equation}
For $T\rightarrow +\infty$, $k_{m}^{2}\rightarrow -k_{0}^{2}$ and $\gamma=-{T\over \xi}k^{2}$.

\section{Particular equations of state}
\label{sec_eos}

The hydrodynamic equations (\ref{intro1})-(\ref{intro3}) or the
generalized Keller-Segel model (\ref{intro4})-(\ref{intro5})
incorporate a pressure force $-\nabla p(\rho)$ associated with a
barotropic equation of state $p(\rho)$.  At equilibrium, the system
 satisfies a relation of the form
\begin{equation}
\label{eos1}
\nabla p=\rho\nabla c.
\end{equation}
This corresponds to a condition of hydrostatic balance between the
pressure force and the chemotactic attraction. For inhomogeneous
systems at equilibrium, the density is a function $\rho=\rho(c)$ of
the concentration of the chemical obtained by integrating
Eq. (\ref{eos1}). We have the identities
\begin{equation}
\label{eos1b}
\int^{\rho}\frac{p'(x)}{x}dx=c, \qquad \frac{p'(\rho)}{\rho}=\frac{1}{\rho'(c)}, \qquad p'(c)=\rho.
\end{equation}
Since $p'(\rho)>0$ in ordinary circumstances, this implies that the
density is an increasing function of the concentration of the chemical,
i.e. $\rho'(c)>0$. On the other hand, for homogeneous systems, the
condition of stability (\ref{intro5ht})  can
be written
\begin{equation}
\label{eos1c}
\frac{p'(\overline{\rho})}{\overline{\rho}}\ge \frac{\lambda}{k_{0}^{2}}.
\end{equation}
This relation determines the range of densities $\overline{\rho}$ for which the
system is stable, depending on the form of the equation of state
$p(\rho)$. In particular, the system is stable whatever the value of the density if $\min_{\rho}\lbrack {p'(\rho)}/{\rho}\rbrack\ge 
\lambda/k_{0}^{2}$ and unstable (for sufficiently large wavelengths) 
whatever the value of the density if $\max_{\rho}\lbrack
{p'(\rho)}/{\rho}\rbrack\le
\lambda/k_{0}^{2}$. The pressure force in Eq. (\ref{intro2}) 
can take into account different effects such as anomalous diffusion or
close packing effects. Typically, three kinds of pressure law
$p(\rho)$ have been considered in the chemotactic literature:

(i) In the standard case, the pressure is a linear function of the density
\begin{equation}
\label{eos2}
p=\rho T.
\end{equation}
This is similar to an isothermal equation of state where $T$ is an effective temperature \cite{sc}. When this law is substituted in the drift-diffusion equation (\ref{intro4}) we recover the standard Keller-Segel model \cite{ks}:
\begin{equation}
{\partial\rho\over\partial t}=\nabla \cdot  \left (D\nabla \rho-\chi\rho\nabla c\right ), \label{eos3}
\end{equation}
where we have introduced the diffusion coefficient $D$ through the
Einstein relation $D=\chi T$. The equilibrium state is the Boltzmann
distribution $\rho=Ae^{\beta c}$. For the pressure law (\ref{eos2}),
the velocity of sound has the constant value $c_{s}^{2}=T$. Therefore,
the stability criterion (\ref{intro5ht}) can be rewritten in the form
\begin{equation}
\label{eos4t}
T\ge T_c\equiv \frac{\lambda \overline{\rho}}{k_0^2}.
\end{equation}
For a fixed density $\overline{\rho}$, it defines a critical temperature $T_{c}$ below which the system is unstable. Alternatively, for a fixed temperature $T$, the stability criterion (\ref{intro5ht}) can be rewritten in the form
\begin{equation}
\label{eos4}
\overline{\rho}\le \overline{\rho}_{crit}\equiv \frac{k_{0}^{2}T}{\lambda}.
\end{equation}
The system becomes unstable above a certain critical density
${\rho}_{crit}$.

\begin{figure}
\centerline{
\psfig{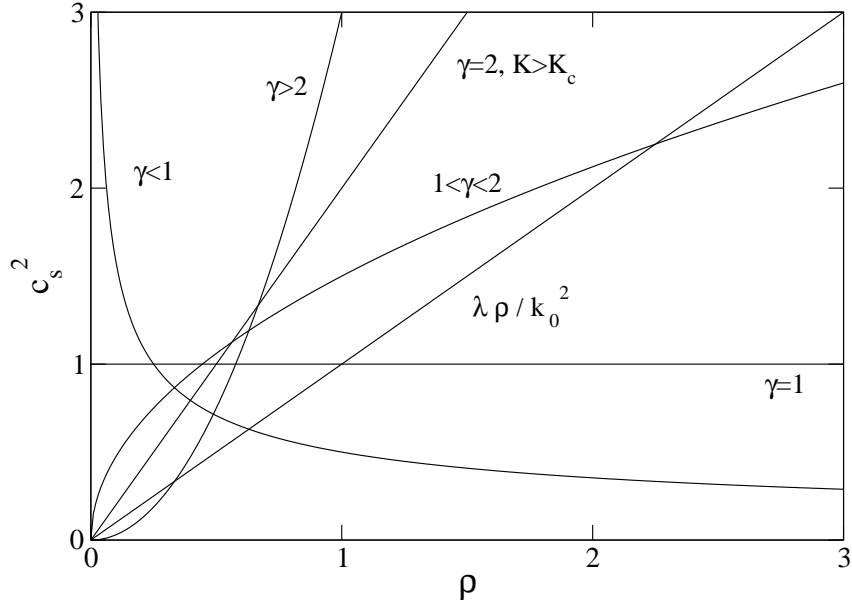}}
\caption{Graphical construction determining the instability threshold (expressed in terms of the density) for a polytropic equation of state with index $\gamma$.  The system is stable if $c_s^2=K\gamma{\rho}^{\gamma-1}\ge \lambda{\rho}/k_{0}^{2}$ and unstable to sufficiently large wavelengths otherwise. We have taken $K=1$ and $\lambda/k_{0}^{2}=1$. } \label{polytropes}
\end{figure}

\begin{figure}
\centerline{
\psfig{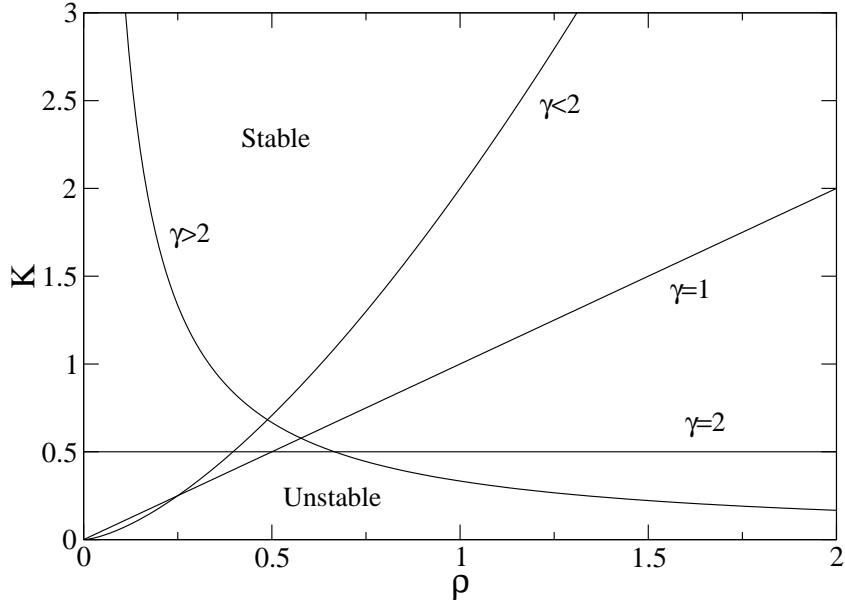}}
\caption{Stability diagram of a spatially uniform polytropic gas with an attractive Yukawa potential of interaction. The curves represent $K_{c}(\rho)$ or $\rho_{crit}(K)$ and separate the stable region (upper region) from the unstable one (lower region). For $\gamma=1$, we recover the isothermal case with $K=T$. We have taken  $\lambda/k_{0}^{2}=1$.  } \label{diagphasepoly}
\end{figure}

(ii) In \cite{lang,cschandra}, we have proposed to take into account anomalous
diffusion by using an equation of state of the form
\begin{equation}
\label{eos5}
p=K\rho^{\gamma}.
\end{equation}
This is similar to a polytropic equation of state where $K$ plays the role of a  polytropic temperature. When this law is substituted in the drift-diffusion equation (\ref{intro4}) we obtain the generalized Keller-Segel model studied in \cite{lang,cschandra}:
\begin{equation}
{\partial\rho\over\partial t}=\nabla \cdot \left\lbrack \chi \left (K\nabla \rho^{\gamma}-\rho\nabla c\right )\right\rbrack. \label{eos6}
\end{equation}
The equilibrium state is the Tsallis distribution $\rho=\lbrack
\lambda+(\gamma-1)/(K\gamma) c\rbrack^{1/(\gamma-1)}$ \cite{tsallis}.
For the pressure law (\ref{eos5}), the velocity of sound has the value
$c_{s}^{2}=K\gamma\rho^{\gamma-1}$.  Therefore,
the stability criterion (\ref{intro5ht}) can be rewritten in the form
\begin{equation}
\label{eos4tt}
K\ge K_c\equiv \frac{\lambda {{\rho}}^{2-\gamma}}{\gamma k_0^2}.
\end{equation}
For a fixed density $\overline{\rho}$, it defines a critical
polytropic temperature $K_{c}$ below which the system is
unstable. Alternatively, for a fixed polytropic temperature $K$, we
can express the stability criterion (\ref{intro5ht}) as a function of
the density. We need to distinguish three cases (see
Figs. \ref{polytropes} and \ref{diagphasepoly}): (a) For $\gamma<2$, the stability criterion
can be written in the form
\begin{equation}
\label{eos7}
\overline{\rho}\le \overline{\rho}_{crit}\equiv \left (\frac{K\gamma k_{0}^{2}}{\lambda}\right )^{\frac{1}{2-\gamma}}.
\end{equation}
The system becomes unstable above a critical density. (b) For $\gamma>2$,
 the stability criterion can be written in the form
\begin{equation}
\label{eos8}
\overline{\rho}\ge \overline{\rho}_{crit}\equiv \left (\frac{\lambda}{K\gamma k_{0}^{2}}\right )^{\frac{1}{\gamma-2}}.
\end{equation}
The system becomes unstable below a critical density. (c) For $\gamma=2$,
 the stability criterion can be written in the form
\begin{equation}
\label{eos9}
K\ge K_{c}\equiv \frac{\lambda}{2k_{0}^{2}}.
\end{equation}
The instability threshold is independent on the density.

\begin{figure}
\centerline{
\psfig{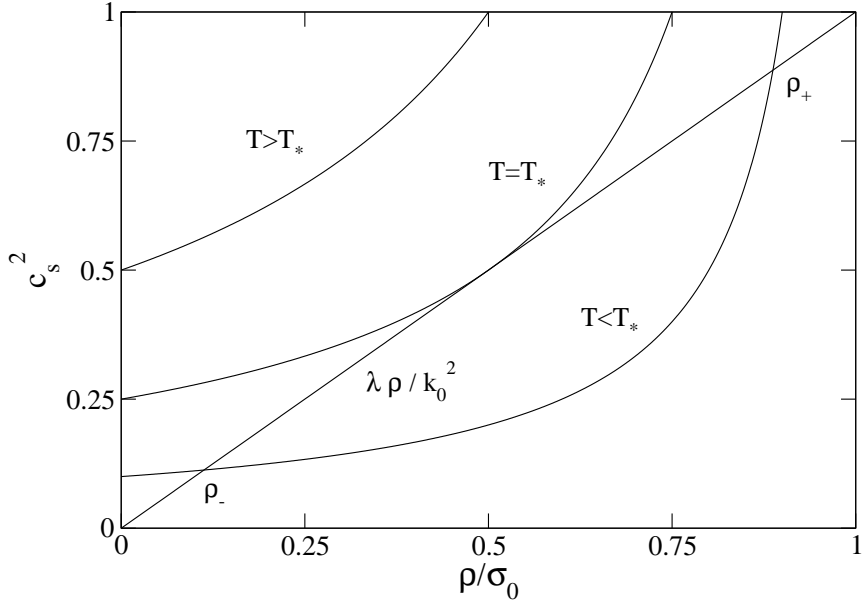}}
\caption{Graphical construction determining the instability threshold (expressed in terms of the density) for an equation  of state of the form (\ref{eos10}) with temperature $T$. The system is stable if $c_s^2=T/(1-\rho/\sigma_0)\ge \lambda{\rho}/k_{0}^{2}$ and unstable to sufficiently large wavelengths otherwise. We have taken $\sigma_{0}=1$ and $\lambda/k_{0}^{2}=1$ leading to a critical temperature $T_{*}=1/4$.  } \label{fermidirac}
\end{figure}

\begin{figure}
\centerline{
\psfig{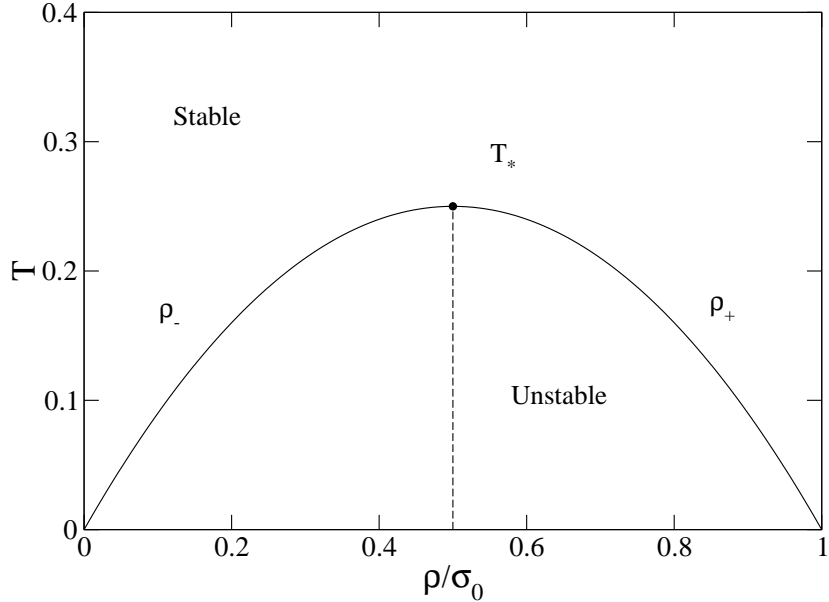}}
\caption{Stability diagram of a spatially uniform  gas described by the pressure law (\ref{eos10}) with an attractive Yukawa potential of interaction. The curve represents $T_{c}(\rho)$ or $\rho_{\pm}(T)$ and separates the stable region from the unstable one. There exists a critical point $T_{*}$ above which the homogeneous phase is always stable whatever the density. Below $T_{*}$, the homogenous phase is stable for $\rho<\rho_{-}$, unstable for $\rho_{-}<\rho<\rho_{+}$ and stable again for $\rho>\rho_{+}$. This corresponds to a reentrant phase.
 We have taken $\sigma_{0}=1$ and $\lambda/k_{0}^{2}=1$ leading to a
 critical temperature $T_{*}=1/4$.  } \label{diagphasefermi}
\end{figure}

(iii) As a result of chemotactic collapse, the standard Keller-Segel
model can lead to finite time singularities and Dirac peaks
\cite{horstmann,sc,post}. In reality, these singularities 
are unphysical because the cells have a finite size and cannot be
compressed indefinitely. Therefore, in more realistic models, we
expect that the pressure $p(\rho)$ tends to zero for low densities
$\rho\rightarrow 0$ and rapidly increases for large densities. This
takes into account the fact that cells do not interpenetrate due to
their finite size and this prevents overcrowding. In \cite{degrad},
one of us has proposed to take into account volume filling and finite
size effects by using an equation of state of the form
\begin{equation}
\label{eos10}
p=-T\sigma_{0}\ln(1-\rho/\sigma_{0}).
\end{equation}
For low densities $\rho\rightarrow 0$, we recover the linear equation of state $p=\rho T$ and for high densities, close to the maximum allowable density $\sigma_{0}$, the pressure rapidly increases and diverges when $\rho\rightarrow \sigma_{0}$. If $a$ represents the
typical size of the cells, we have $\sigma_{0}\sim 1/a^{d}$ where $d$
is the dimension of space. When this law is substituted in the drift-diffusion equation (\ref{intro4}) we obtain the generalized Keller-Segel model studied in \cite{degrad}:
\begin{equation}
{\partial\rho\over\partial t}=\nabla \cdot \left\lbrack \chi \left (\frac{T}{1-\rho/\sigma_{0}}\nabla \rho-\rho\nabla c\right )\right\rbrack. \label{eos11}
\end{equation}
The steady state of this equation is a Fermi-Dirac distribution in
physical space $\rho=\sigma_{0}/(1+\lambda e^{-\beta c})$ putting an
upper bound on the density: $\rho\le \sigma_{0}$ \footnote{Equation
(\ref{eos11}) can be viewed as a generalized mean field Fokker-Planck equation
\cite{gen} with a constant mobility and a nonlinear diffusion. A
related model, corresponding to a constant diffusion and a variable
mobility $\chi(\rho)=\chi(1-\rho/\sigma_{0})$ vanishing above the
close packing value $\sigma_{0}$, has been considered in
\cite{hillen,gen,degrad}. The two models have the same steady states and are
associated with the same free energy. They present therefore the same
general properties. The details of the evolution may, however, be
different in the two models.}.  For the pressure law (\ref{eos10}),
the velocity of sound is $c_{s}^{2}=T/(1-{\rho}/\sigma_{0})$. Therefore,
the stability criterion (\ref{intro5ht}) can be rewritten in the form
\begin{equation}
\label{eos4ttt}
T\ge T_c\equiv \frac{\lambda \overline{\rho} (1-\overline{\rho}/\sigma_0)}{k_0^2}.
\end{equation}
For a fixed density $\overline{\rho}$, it defines a critical
temperature $T_{c}$ below which the system is
unstable. Alternatively, for a fixed temperature $T$, we
can express the stability criterion (\ref{intro5ht}) as a function of
the density. The system is stable if
\begin{equation}
\overline{\rho}^{2}-\sigma_{0}\overline{\rho}+\frac{T\sigma_{0}k_{0}^{2}}{\lambda}\ge 0, \label{eos12}
\end{equation}
and unstable to large wavelengths otherwise.  The discriminant of this
equation is $\Delta=\sigma_{0}^{2}-4T\sigma_{0}k_{0}^{2}/\lambda$. If
$\Delta<0$, corresponding to
\begin{equation}
T>T_{*}\equiv \frac{\lambda \sigma_{0}}{4 k_{0}^{2}}, \label{eos13}
\end{equation}
the system is stable whatever the value of the density (see
Figs. \ref{fermidirac} and \ref{diagphasefermi}). Alternatively, if $T<T_{*}$, the system is
stable for $\overline{\rho}<\overline{\rho}_{-}$ and
$\overline{\rho}_{+}<\overline{\rho}<\sigma_{0}$ and unstable for
$\overline{\rho}_{-}<\overline{\rho}<\overline{\rho}_{+}$ where
\begin{equation}
\overline{\rho}_{\pm}=\frac{\sigma_{0}}{2}\left (1\pm\sqrt{1-\frac{T}{T_{*}}}\right ). \label{eos14}
\end{equation}
For low densities ($\rho\ll\sigma_{0}$), the homogenous phase is
stable because the chemotactic attraction is not strong enough to
overcome diffusive effects (like for a low density isothermal gas (i))
and for high densities ($\rho\rightarrow \sigma_{0}$), the homogeneous
phase is stabilized by pressure effects due to close packing.
Other generalized Keller-Segel models of chemotaxis are discussed in
\cite{nfp} in relation with nonlinear mean field Fokker-Planck
equations.

\section{Conclusion}
\label{sec_conc}

In this paper, we have studied the chemotactic instability of an
infinite and homogeneous distribution of cells whose dynamics is
described by the hydrodynamic equations
(\ref{intro1})-(\ref{intro3}). We have shown the analogy with the
classical Jeans instability in astrophysics. This close analogy
between two systems of a very different nature (stars and bacteria) is
very intriguing and deserves to be developed and emphasized. As is
well-known, an infinite and homogeneous distribution of stars is not a
stationary state of the gravitational Euler-Poisson system \cite{bt}.
However, if we make the Jeans swindle (which is made in any textbook
of astrophysics), the equations for the linear perturbations are the
same as in the biological problem (\ref{intro1})-(\ref{intro3}) when
$\xi=k_0=0$. Therefore, the two systems are really analogous. The main
differences between the chemotactic problem and the Jeans problem are
due to the presence of (i) a friction force $-\xi {\bf u}$ in the
Euler equation (\ref{intro2}) and (ii) a shielding length $k_0^{-1}$
in the equation (\ref{intro3}) determining the potential of
interaction (played here by the concentration of the secreted
chemical). We have studied the effect of these terms in detail. This
leads to a generalization of the Jeans instability analysis. The
shielding length determines a critical velocity of sound
$(c_{s})_{crit}=({\lambda\overline{\rho}}/{k_{0}^{2}} )^{1/2}$, so
that the system is always stable if $c_s>(c_{s})_{crit}$ and becomes
unstable to large wavelengths if $c_s<(c_{s})_{crit}$. Therefore, the
system experiences a phase transition from a homogeneous distribution
to an inhomogeneous distribution when the velocity of sound passes
below a critical value $(c_{s})_{crit}$. In the usual Jeans problem,
the system is always unstable to large wavelengths since
$(c_{s})_{crit}=+\infty$. The condition of instability corresponds to
$k\le k_{J}$ where $k_{J}=(\lambda\overline{\rho}/c_{s}^{2})^{1/2}$ is
the Jeans wavenumber.  For $k_{0}\neq 0$ and $c_s<(c_{s})_{crit}$, the
condition of instability is $k\le k_{m}\equiv
\sqrt{k_{J}^{2}-k_{0}^{2}}$. Therefore, the effect of the shielding is to shift
the instability to larger wavelengths with respect to the Jeans
length. In order to measure the influence of the friction parameter
$\xi$, we have introduced a wavenumber $k_d=\xi/(2c_s)$ and a
dimensionless number $F=(k_d/k_J)^2$. The square root of this number
$F^{1/2}\sim \xi t_D$ corresponds to the ratio between the dynamical
time $t_D\sim 1/\sqrt{\overline{\rho}\lambda}$ and the friction time
$\xi^{-1}$.  For $F=0$, Eq. (\ref{intro2}) reduces to the Euler
equation describing a purely inertial evolution ($t_D\ll \xi^{-1}$)
and for $F\rightarrow +\infty$, Eqs. (\ref{intro1})- (\ref{intro2})
lead to the generalized Smoluchowski equation (\ref{intro4})
describing an overdamped evolution ($\xi^{-1}\ll t_D$). We have
introduced a wavenumber $k_c(F)$ which separates, in the zone of
stable wavenumbers ($k>k_m$), the case of purely exponential decay
($k_m<k<k_c$) from the case of damped oscillations ($k>k_c$). We have
also determined, in the unstable zone ($k<k_m$), the wavenumber $k_*$
corresponding to the maximum growth rate. In the Newtonian model (no
shielding) it is equal to $k_*=0$ corresponding to infinite
wavelengths. In the Yukawa model, it corresponds to a finite
wavelength given by Eq. (\ref{yukawa24}), independent on the friction
parameter $\xi$. For $c_s\rightarrow (c_s)_{crit}$, $k_*\rightarrow 0$
and for $c_s\rightarrow 0$, $k_*\rightarrow +\infty$ corresponding to
small wavelengths. We have found that the shielding length present in
the chemotactic model solves many problems inherent to the Jeans
analysis. Indeed, there is {\it no Jeans swindle} in the chemotactic
problem and the maximum growth rate occurs for a {\it finite}
wavelength (when $k_{0}\neq 0$) instead of an infinite wavelength
(when $k_{0}=0$). Therefore, the mathematical problem of linear
dynamical stability is better posed in biology than in astrophysics
since it avoids the Jeans swindle.

The linear stability analysis performed in this paper gives the
condition of instability (in Sec. \ref{sec_eos}, we have expressed
this condition of instability in terms of the density for different
equations of state $p(\rho)$ used in the literature) and describes the
early development of the instability.  When the condition of
instability (\ref{intro7ht}) is fulfilled, the perturbation grows
until the system can no longer be described by equilibrium or near
equilibrium equations. Therefore, the next step is to study the
chemotactic collapse in the nonlinear regime to see the formation of
patterns like clusters and filaments.  A large number of studies in
applied mathematics (see the extensive list of references given in the
review
\cite{horstmann}) and physics
\cite{sc,lang,post,degrad} have considered the overdamped limit of the
model (\ref{intro1})-(\ref{intro3}) leading to the Keller-Segel model
(\ref{intro4})-(\ref{intro5}), similar to the Smoluchowski-Poisson system (\ref{intro9})-(\ref{intro10}). For this parabolic
model, chemotactic collapse leads to the formation of round
clusters. The evolution of an individual cluster in the nonlinear
regime can be studied by considering spherically symmetric solutions
of the Keller-Segel model. The standard Keller-Segel model
(\ref{eos3}) leads to the formation of Dirac peaks (for $d\ge 2$)
\cite{horstmann,sc,post}. In the regularized model (\ref{eos11}), the
Dirac peaks are replaced by smooth aggregates
\cite{hillen,degrad}. These aggregates interact with each other and
lead to a coarsening process where the number of clusters $N(t)$
decays with time as they collapse to each other. This process may
share some analogies with the aggregation of vortices in
two-dimensional turbulence \cite{turb}. We expect that the decay of
the number of clusters depends on the effective range of interaction
mediated by the chemical $k_0^{-1}$ (shielding length). If the
shielding length is small the clusters do not ``see'' each other and
the decay of $N(t)$ should be slowed down or even stopped. In that
case, we obtain a quasi stationary state made of clusters separated
from each other by a distance of the order of the shielding length
$k_0^{-1}$. If we take into account inertial effects, using the
hyperbolic model (\ref{intro1})-(\ref{intro3}) instead of the
parabolic model (\ref{intro4})-(\ref{intro5}), the collection of
isolated clusters is replaced by a network pattern with nodes
(clusters) separated by chords \cite{gamba,filbet,csbio}. The
filaments between two nodes have a length of the order of
$k_0^{-1}$. Again, the number of nodes should decay with
time. However, if the shielding length $k_0^{-1}$ is small, the
evolution is slowed down and we get a quasi equilibrium state with a
filamentary structure corresponding to the initiation of a vasculature
\cite{gamba}.




\end{document}